\documentclass[modern]{aastex63}

\newcommand{\PIP}{P_{\rm{IP}}}
\newcommand{\Prot}{P_{\rm{rot}}}

\received{}
\revised{}
\accepted{}
\submitjournal{ApJL}

\shorttitle{}
\shortauthors{Reinhold et al.}

\begin{document}

\title{Measuring periods in aperiodic light curves -- Applying the GPS method to infer rotation periods of solar-like stars}

\correspondingauthor{Timo Reinhold}
\email{reinhold@mps.mpg.de}

\author[0000-0002-1299-1994]{Timo Reinhold}
\affiliation{Max-Planck-Institut f\"ur Sonnensystemforschung, Justus-von-Liebig-Weg 3, 37077 G\"ottingen, Germany}

\author[0000-0002-8842-5403]{Alexander I. Shapiro}
\affiliation{Max-Planck-Institut f\"ur Sonnensystemforschung, Justus-von-Liebig-Weg 3, 37077 G\"ottingen, Germany}

\author[0000-0002-3418-8449]{Sami K. Solanki}
\affiliation{Max-Planck-Institut f\"ur Sonnensystemforschung, Justus-von-Liebig-Weg 3, 37077 G\"ottingen, Germany}

\author[0000-0001-9820-8464]{Gibor Basri}
\affiliation{Department of Astronomy, University of California, Berkeley, CA 94720, USA}




\begin{abstract}
Light curves of solar-like stars are known to show highly irregular variability. As a consequence, standard frequency analysis methods often fail to detect the correct rotation period. Recently, \citet{GPS_I} showed that the periods of such stars could still be measured by considering the Gradient of the Power Spectrum (GPS) instead of the power spectrum itself. In this study, the GPS method is applied to model light curves of solar-like stars covering all possible inclination angles and a large range of metallicities and observational noise levels. The model parameters are chosen such that they resemble those of many stars in the Kepler field. We show that the GPS method is able to detect the correct rotation period in $\approx 40\%$ of all considered cases, which is more than ten times higher than the detection rate of standard techniques. Thus, we conclude that the GPS method is ideally suited to measure periods of those Kepler stars lacking such a measurement so far. We also show that the GPS method is significantly superior to auto-correlation methods when starspot lifetimes are shorter than a few rotation periods. GPS begins to yield rotation periods that are too short when dominant spot lifetimes are shorter than one rotation period. We conclude that new methods are generally needed to reliably detect rotation periods from sufficiently aperiodic time series --- these periods will otherwise remain undetected.
\end{abstract}



\section{Introduction}\label{sec:intro}
Stellar rotation periods are usually measured by searching for periodic patterns in long-term (photometric) time series caused by magnetic features transiting the stellar disk several times. The high-precision, long-term photometry of the Kepler space telescope has proven to be an ideal data set for large rotation period surveys (e.g., \citealt{Reinhold2013,McQuillan2014,Santos2019}, etc.). In particular, these surveys have shown that rotation periods of stars with regular light curve patterns can reliably be detected by standard frequency analysis methods such as the Lomb-Scargle periodogram \citep{Reinhold2013,Reinhold2015}, the auto-correlation function \citep{McQuillan2014}, or wavelet power spectra \citep{Santos2019,Santos2021}.

These methods, however, reach their limits when applied to stars around solar spectral type that show comparably small and irregular variability. The Sun also shows an irregular light curve pattern, which is mostly caused by the relatively short lifetimes of the sunspots (days to weeks, see \citealt{Petrovay1997}) compared to the solar rotation period of roughly 27 days. A similar behavior is seen for stars around solar spectral type, which often show highly irregular light curves. As a consequence, rotation periods could only be detected for 16\% of the G-type dwarfs in the Kepler field (see Table~1 and 2 in \citealt{McQuillan2014}). \citet{Basri2022} showed that the stars they were unable to find reliable periods for are the ones whose starspot lifetimes are shorter than a couple of rotation periods.

Recently, \citet{Reinhold2021} tested the performance of the auto-correlation function (ACF) on simulated light curves with solar variability patterns. These authors showed that the ACF is not well-suited for the period detection in such aperiodic stars because the method (by design) searches for a repeating variability pattern. \citet{Reinhold2021} further showed that the period detection rate is as low as $\approx 3\%$ when setting similar constraints on the auto-correlation peak heights as used in \citet{McQuillan2014}.

Thus, measuring periodic signals in a non- or hardly periodic time series is challenging and requires novel techniques. Recently, \citet{GPS_I} showed that the periods of such stars could still be measured by considering the Gradient of the (global wavelet) Power Spectrum (GPS) instead of the power spectrum itself. This so-called GPS method is able to detect the correct rotation period in an aperiodic time series. In particular, it was shown that the GPS method is able to determine the correct rotation period of the Sun \citep{GPS_II}. Furthermore, the GPS method has been successfully applied to Kepler stars with previously determined rotation periods \citep{GPS_III}.

In contrast to classical period analysis methods (such as the Lomb-Scargle periodogram, auto-correlation functions, or power spectra), the GPS method does not require a repeatable pattern in the time series. Instead, the GPS method is mostly sensitive to the shape and in particular the width of the light curve profile of individual starspot transits averaged over time (see Sect.~\ref{lifetime}). Consequently, the GPS method even works in cases when the magnetic features live less long than the stellar rotation period such that no reoccurring transits of the same magnetic features are needed. These characteristics makes it a powerful tool for measuring periods in aperiodic time series.

Here, we test the GPS method on simulated light curves with irregular variability, and show that it reaches a detection rate of $\approx 40\%$, which is more than ten times higher than the detection rates obtained by classical methods (such as auto-correlation functions). In a forthcoming publication, we will apply this method to all Kepler stars with near-solar effective temperatures with the goal of measuring periods for the large sample of stars lacking this information so far.

\section{Data \& Methods}

\subsection{The model light curves}\label{models}
In this study, we primarily apply the GPS method to simulated light curves with solar-like variability. The light curves rely on models of solar brightness variability, SATIRE (Spectral And Total Irradiance REconstruction, \citealt{Fligge2000, Krivova2003,Solanki2013}). These models build on the solar paradigm to simulate light curves with solar effective temperature, rotation period, and activity level, and have recently been extended to modeling stars observed at different inclination angles \citep{Nemec2020,Nemec2020b} and metallicities \citep{Witzke2020,Reinhold2021}.

The models considered here span 61 years of solar data between March 1948 and May 2009, i.e., the data almost cover 6 full solar activity cycles with epochs of high and low activity. The models were calculated for ten inclination angles between pole-on ($i=0^\circ$) and edge-on ($i=90^\circ$) views, and nine different metallicities $\rm -0.4 \leq [Fe/H] \leq 0.4$~dex (in steps of 0.1~dex). In particular, the rotation period of all model light curves is equal to the solar one, which roughly equals 27~days as seen from Earth.

To be comparable to Kepler observations later on, we chose a Monte Carlo approach to account for a wide range of possible stellar parameters such as inclination angles, metallicities, and noise levels. Following the strategy described in \citet{Reinhold2021}, the input model parameters inclination and metallicity are chosen such that they resemble the distribution in the Kepler field. From these distributions, we chose a random combination of inclination and metallicity, and pick the model closest to the chosen values. From this model, we randomly pick a 4-yr segment from the whole 61-years time series, and remove the long-term variability of the activity cycle (see \citealt{Reinhold2021} for details). The GPS method is then applied to these \textit{Keplerized} light curves as described in the following.

Since the light curves are noise-free, we add white noise with zero mean and standard deviation $\sigma$ to the data. The noise level $\sigma$ is determined by the apparent magnitude of the observed star. We adopt the distribution of apparent magnitudes (Kp) of solar-like stars in the Kepler field, and compute different noise realizations following the procedure described in \citet{Reinhold2020}. In total, we have considered 5000 samples each for the noise-free and the noisy light curves. For more details, we refer the reader to \citet{Reinhold2021}, who used the exact same light curves as we do.

In the following, the GPS method will primarily be applied to the physics-based SATIRE model light curves. In order to directly test the effect of starspot lifetimes on the efficacy of the GPS and auto-correlation methods, we also utilized some of the models described in \citealt{Basri2020} (see Sect.~\ref{lifetime}). These do not necessarily mimic stars similar to the Sun, but instead were computed to include the general behavior of light curves of Kepler stars whose periods have been determined. The models do not simulate physical flux emergence processes, instead sampling light curves produced by many randomized manifestations of starspots whose general parameters are specified and varied. In particular, a set of models was employed with an average of six spots (all with the same maximum size) present at a time, distributed randomly over the stellar sphere, and viewed at an inclination of 60 degrees. Starspot lifetimes for these noise-free model sets were set at values of 0.25, 0.5, 1, 2, 3, and 4 rotation periods. Each model set contains 1000 different instances of random spot placements and birth dates and each run has 50 rotations. To better illustrate the effect of different spot lifetimes, we refrain from adding noise to these models.

\subsection{The GPS method}

\begin{figure}
  \centering
  \includegraphics[width=\textwidth]{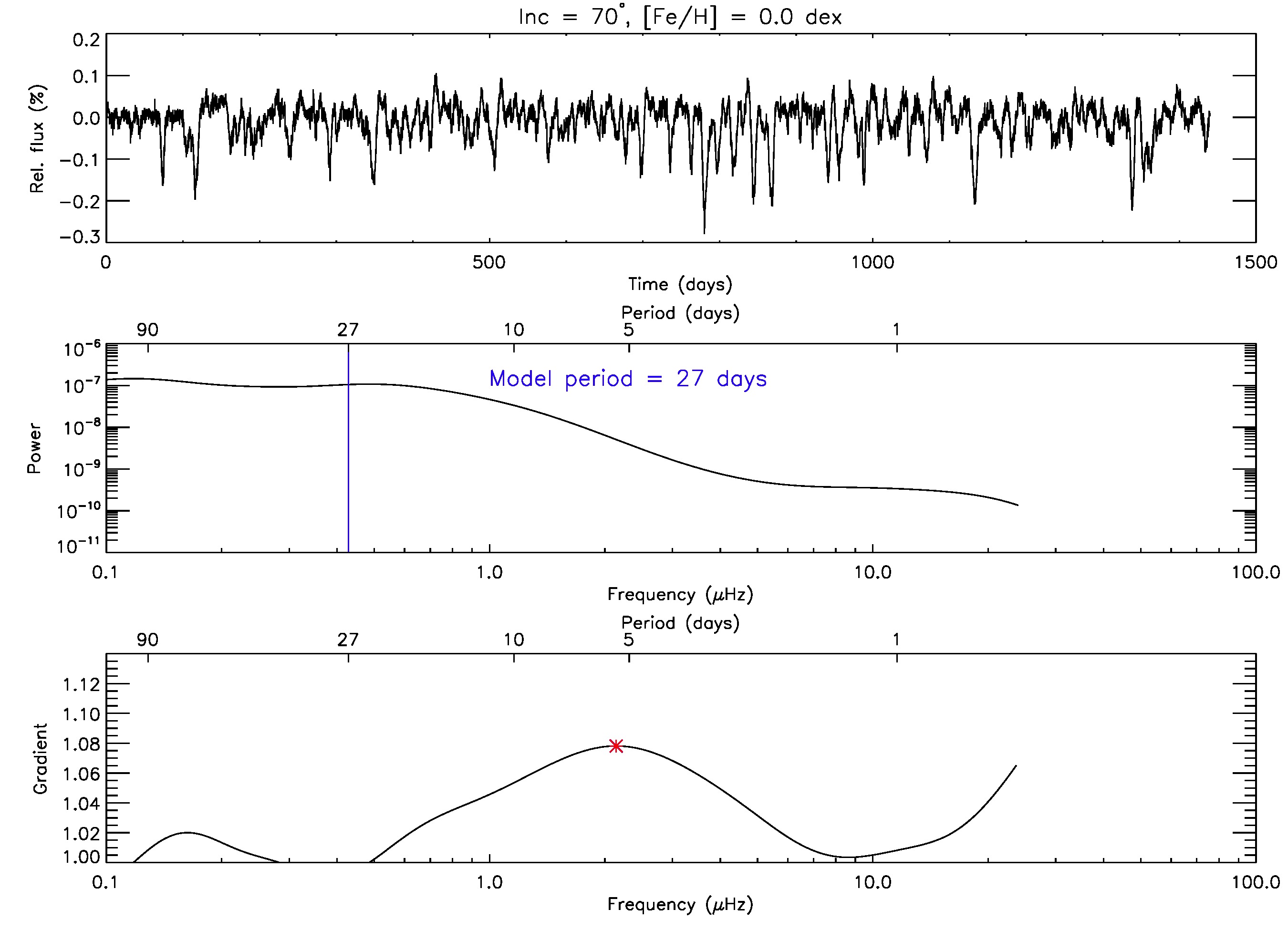}
  \caption{Upper panel: \textit{Keplerized} model light curve (including noise) with an inclination of $i=70^\circ$ and solar metallicity $[Fe/H]=0.0$ dex. Individual spot transits are visible but there is no obvious periodicity. Middle panel: global wavelet power spectrum. The vertical blue line indicates the model rotation period of 27 days. The power spectrum does not show any peak at this period. Lower panel: Gradient of the Power Spectrum (GPS). The highest peak indicates the period at the inflection point $\PIP=5.41$ days. This period can be used to infer the rotation period.}
  \label{lc_example}
\end{figure}

Fig.~\ref{lc_example} shows an example from our set of (noisy) model light curves (top panel). Individual spot transits are clearly visible but without any obvious periodicity. This observation is confirmed by the global wavelet\footnote{Here, we use a 6th order Paul wavelet (e.g., \citealp{Torrence1998}).} power spectrum of the light curve (middle panel).  The power spectrum increases toward longer periods but does not show a distinct peak. Instead, it shows a plateau shape at the model rotation period of 27~days. Thus, the rotation period would not have been detected by this method.

Instead of searching for peaks of the power spectrum, we compute the Gradient of the Power Spectrum (GPS), which is shown in the lower panel of Fig.~\ref{lc_example}. The maximum of the GPS corresponds to the position of the inflection point (IP), i.e., the point where the curvature of the high-frequency tail of the power spectrum changes its sign. In this example, the highest peak is found at the inflection point period $P_{\rm IP} \approx 5.41$ days. \citet{GPS_I} showed that this period can be used to infer the correct rotation period by the simple equation
\begin{equation}\label{eq1}
    \Prot = P_{\rm IP}/\alpha,
\end{equation}
where $\alpha$ is a calibration factor. We will see below how $\alpha$ depends on the model parameters, and what the best choice of $\alpha$ might be for real observations where physical parameters cannot be controlled. For the model in Fig.~\ref{lc_example}, we find that $\alpha=5.41/27=0.20$ is the best choice. The main idea behind the GPS method is that the high-frequency tail of the power spectrum is much less affected by the evolution of magnetic features than the power spectrum peak associated with the rotation period (see Fig.~3 in \citealt{GPS_I}). We note that the inflection point itself does not have a physical meaning, as stated in \citet{GPS_I}. However, we ascribe a sort of physical meaning to the inflection point, which can be associated with the typical dip duration in the light curves caused by active regions crossing the visible disc (see Sect.~\ref{lifetime}).

For some models, the GPS shows more than one peak, i.e., more than one inflection point. From visual inspection, we found that this mostly occurs for cases where the light curve shows some quiet and some active segment. Such models usually return two peaks in the gradient, with one period much smaller and the other peak much larger than the "correct" inflection period. Thus, we discard all cases where more than one GPS peak was found. This criterion removes $\approx25\%$ of the noise-free cases but only $\approx10\%$ of the noisy models. The reason why the noisy models less often show two peaks is that the noise partly washes out the impact of small spot transits that have very short lifetimes (often less than one day). Consequently, more power of the GPS peak can be associated with the profiles of the remaining larger spots.

\section{Results}
\subsection{The calibration factor $\alpha$}\label{alpha}
\begin{figure}
  \centering
  \includegraphics[width=\textwidth]{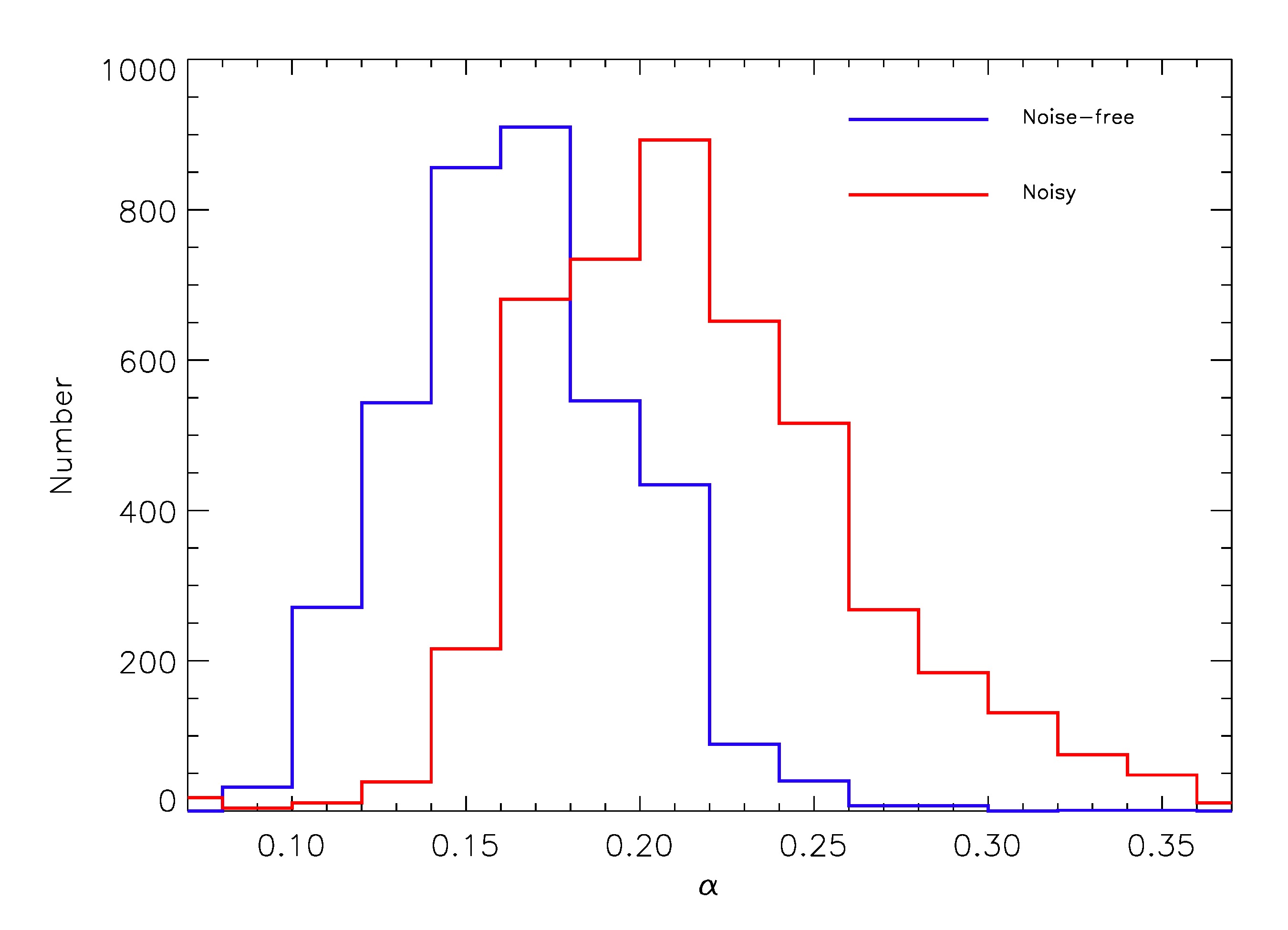}
  \caption{Distribution of the measured inflection point period divided by the model rotation period of 27 days (i.e. $\alpha$) for the noise-free (blue) and the noisy (red) models for 5000 random samples. Adding noise to the light curves shifts the inflection point toward longer periods (i.e. larger values of $\alpha$).}
  \label{alpha_dist}
\end{figure}

To eventually determine the rotation period, we need to know the calibration factor $\alpha$. According to eq.~\ref{eq1}, this factor equals the measured inflection period divided by the model rotation period of 27~days. Following our Monte Carlo approach, we draw random samples of model light curves, and measure the inflection point periods. Fig.~\ref{alpha_dist} shows the distribution of $\alpha$ for the noise-free (blue) and the noisy (red) models. The noise-free distribution has a roughly Gaussian shape with a mean of $\langle\alpha\rangle=0.165$ and standard deviation $\sigma_{\alpha}=0.032$. The distribution for the noisy models is only approximately Gaussian with a mean of $\langle\alpha\rangle=0.217$ and standard deviation $\sigma_{\alpha}=0.047$. Generally, one can see that the noise shifts the distribution to higher $\alpha$ values. We will see below (see Fig.~\ref{alpha_activity}) that the noise-dominated models are mostly responsible for the long tail toward larger $\alpha$ values (i.e. longer inflection periods). We like to emphasize that the specific value of $\alpha$ is independent of the model rotation period. In the following, we show how $\alpha$ depends on different model parameters, and what value of $\alpha$ should be used for real observations (see Sect.~\ref{detection_rate}).

\subsection{Dependence on model parameters}
\begin{figure}
  \centering
  \includegraphics[width=\textwidth]{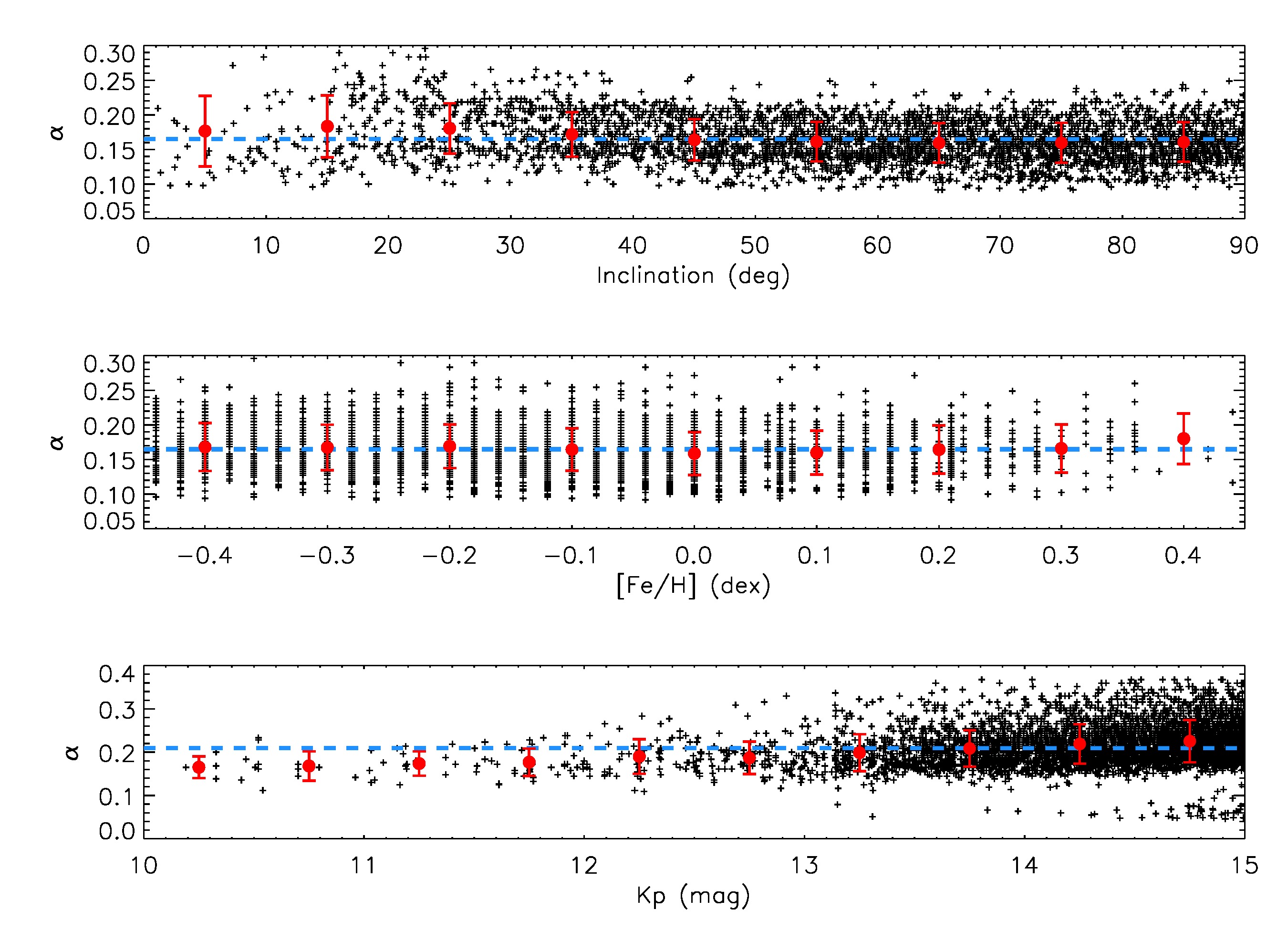}
  \caption{Dependence of $\alpha$ on inclination (first panel), metallicity (second panel), and Kepler magnitude Kp (third panel) for 5000 samples. The dashed blue line indicates the median $\alpha$ of all models, and the red dots and error bars indicate the mean and standard deviation of $\alpha$ in the considered bins. We note that the upper and middle panels show the noise-free models, whereas the lower panel shows the results for the noisy model light curves.}
  \label{alpha_parameter}
\end{figure}

In Fig.~\ref{alpha_parameter}, we can study the dependence of $\alpha$ on the model parameters in more detail. The upper and middle panels show the noise-free models to study the effects of inclination and metallicity independently of the noise, whereas the lower panel shows the $\alpha$ dependence on apparent magnitude (i.e., the effect of noise in the light curves). In all panels, the dashed blue line shows $\langle\alpha\rangle$ and the red dots with error bars indicate the mean and standard deviation in the chosen bins. 

The upper panel shows the dependence of $\alpha$ on inclination for all different metallicities drawn from the input distribution. One can see that $\alpha$ does not show much dependence on inclination down to $i=40^\circ$. Below that angle, the contribution of faculae to the overall variability increases while the spot contribution shrinks (e.g., \citealt{Knaack2001}). The faculae are well visible near the limb and leave a more sinusoidal variability pattern in the light curves than the spots, leading to greater $\alpha$ values eventually. The same is true for high latitude spots, which are seen much longer than spots at lower latitudes. This interplay eventually leads to a bimodal distribution of $\alpha$ for inclinations below $i=30^\circ$.

The middle panel shows that $\alpha$ does not depend on metallicity. As shown in \citet{Witzke2020,Reinhold2021}, a change in metallicity mostly affects the facular contrasts, whereas the spots contrasts are almost unchanged. This result is consistent with the observation that the GPS method is most sensitive to the spot transit profiles.

The strongest dependence of $\alpha$ is found for observational noise such that $\alpha$ strongly increases toward fainter stars. Since the majority (more than 70\%) of the (solar-like) stars in the Kepler field (and thus in our models) are fainter than 14th magnitude, the noise is the dominant uncertainty for $\alpha$.

\begin{figure}
  \centering
  \includegraphics[width=\textwidth]{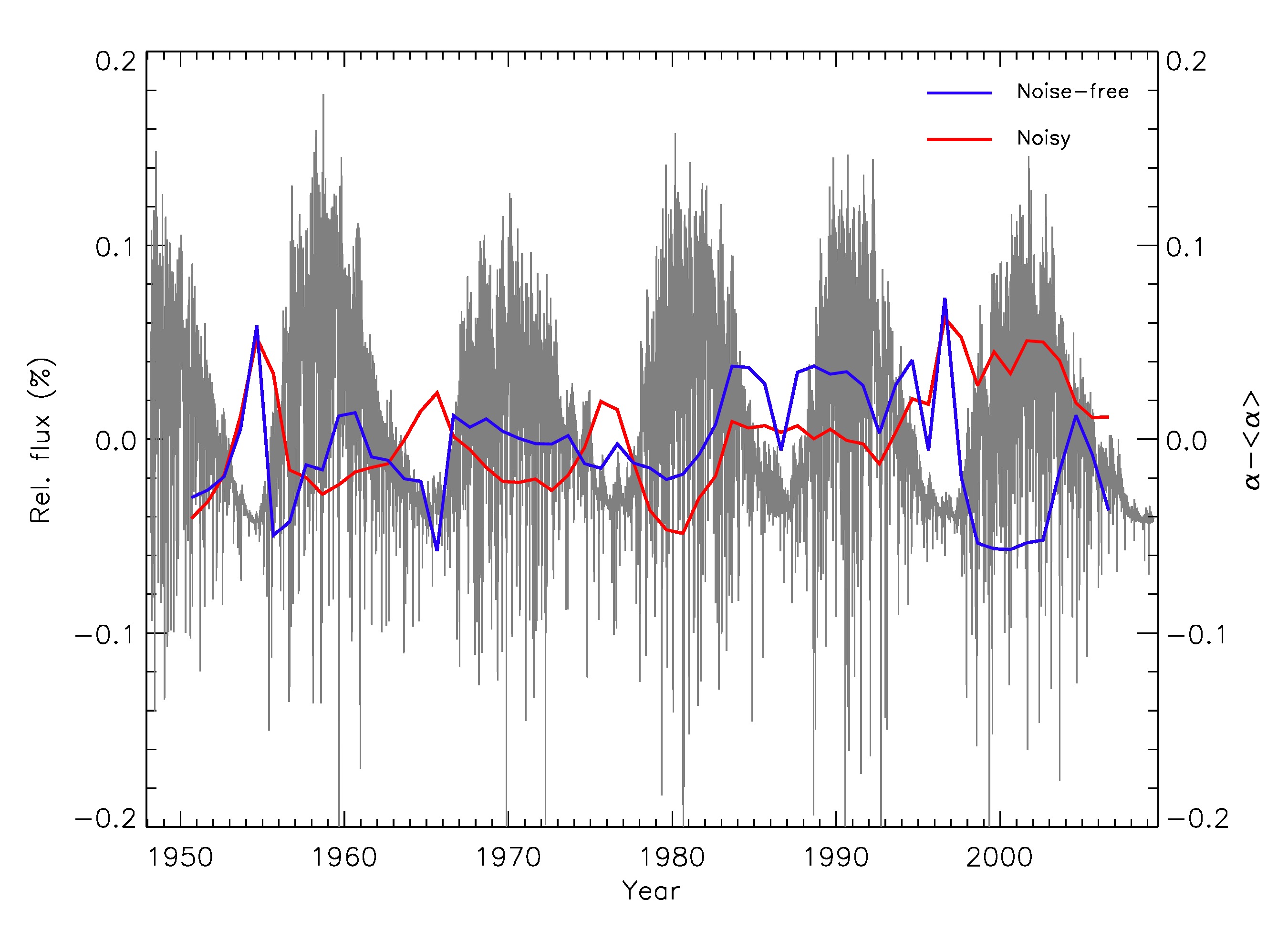}
  \caption{Dependence of $\alpha$ on the activity cycle. The gray data show the underlying model light curve (with $90^\circ$ inclination and solar metallicity) from which we picked 4-yr segments, computed the power spectrum, its gradient and finally the inflection point period (i.e. $\alpha$). The blue (noise-free) and red (noisy) curves shows the yearly averages of $\alpha-\langle\alpha\rangle$ for 5000 randomly drawn samples. We subtracted the mean value $\langle\alpha\rangle$ here to show the overlay with the activity cycle.}
  \label{alpha_activity}
\end{figure}

Additionally to the model parameters, $\alpha$ shows some dependence on the phase of the activity cycle. Fig.~\ref{alpha_activity} shows the 61 years of the undetrended model time series for the solar case, i.e., $i=90^\circ$ and $[Fe/H]=0$ (gray). Since we draw 4-year time series from the whole observing period, the inflection periods are computed for different epochs of the solar cycle. During activity maximum, the (rotational) variability is dominated by spots, whereas faculae are the predominant type of activity during its minima. Thus, the measured $\alpha$ values should depend on the phase of the cycle. This is shown by the blue (noise-free) and red (noisy) lines, which depict the yearly averages of $\alpha$ for all model realizations. To show both the light curve and the $\alpha$ values on top of each other, the mean of the $\alpha$ distribution has been subtracted.

The red curve clearly peaks during activity minima. During these epochs the noise dominates the light curve, which causes the very large $\alpha$ values. During activity maxima, $\alpha$ lies mostly slightly below the average (except for the last cycle). The $\alpha$ distribution of the noise-free models also changes during the modeled period. However, this dependence is not completely clear.

\subsection{Detection rate}\label{detection_rate}
We define the detection rate of a correct period detection as the number of periods in the interval between 24--30 days divided by the number of samples with a single inflection point. The question remains what value of $\alpha$ would be best for measuring the rotation periods, especially later in real data. One choice would be to simply use the mean or median value for all stars. For the noisy sample, we chose the median $\langle\alpha\rangle=0.209$ because the $\alpha$ distribution is not fully Gaussian. Another option would be to use a magnitude-corrected value of $\alpha$. However, the functional dependence of $\alpha$ on Kepler magnitude is not entirely clear (see bottom panel in Fig.~\ref{alpha_parameter}). First, we tried the simplest dependence, namely linear. We also tested the robustness of the resulting period distribution and considered an exponential fit, as the true dependence is not known. Eventually, we found that the resulting rotation period distributions both are very similar. For the fixed value $\alpha=0.209$, we find a detection rate of $39.7\pm0.6\%$. For the linear and exponential corrected $\alpha$ values instead, we find slightly higher detection rates of $41.0\pm0.6\%$ and $41.2\pm0.6\%$, respectively.

Given the fact that the same set of light curves has been analyzed in \citet{Reinhold2021}, this result can directly be compared to the detection rates of auto-correlation functions. We find that the GPS method clearly outperforms the auto-correlation functions, which yield very low detection rates from $17\%$ for a local peak height (LPH) threshold of $LPH>0.1$ down to $1\%$ for stricter criteria as $LPH>0.4$.

\subsection{Effects of Spot Evolution}\label{lifetime}
\begin{figure}
  \centering
  \includegraphics[width=\textwidth]{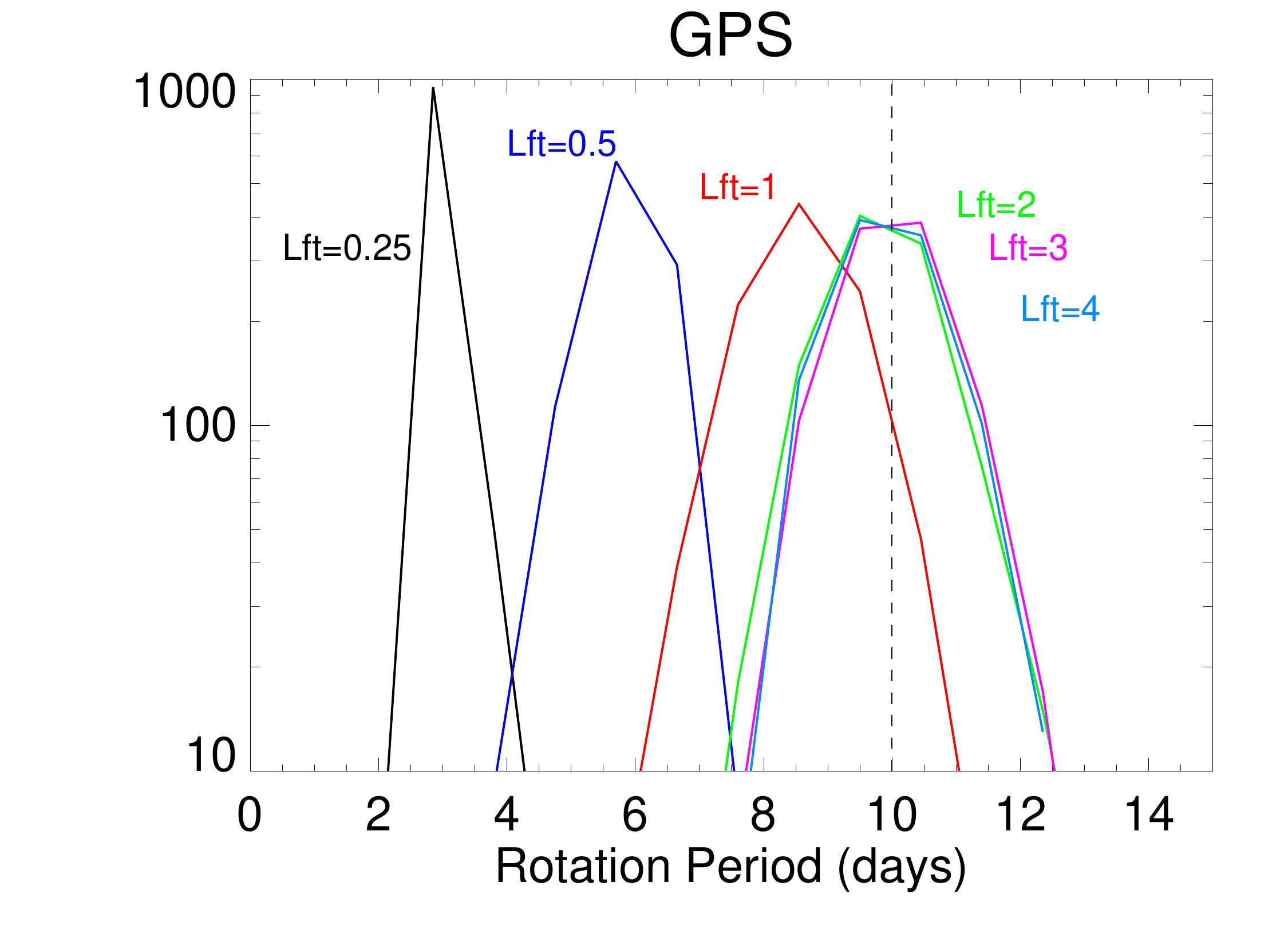}
  \includegraphics[width=\textwidth]{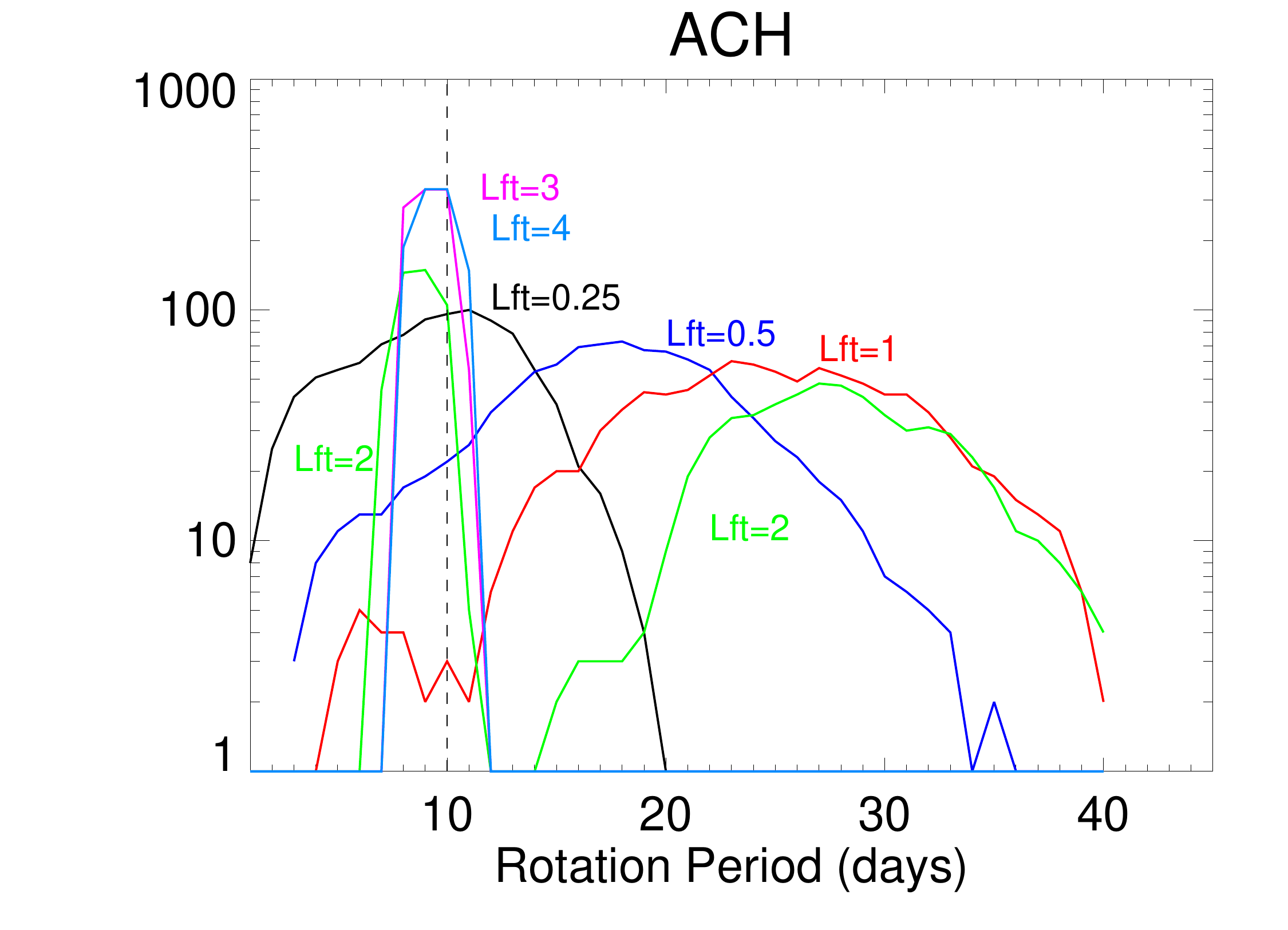}
  \caption{Dependence of derived rotation periods on starspot lifetimes. Upper panel: histograms of derived rotation periods for a model set with different lifetimes utilizing the GPS method. Lower panel: the same but with periods derived using the auto-correlation method. The vertical dashed line in both panels indicates the model rotation period.}
  \label{lifetimes}
\end{figure}

We now directly examine the effect of starspot lifetime on the efficacy of the two methods for finding rotation periods. In our primary physical models spots have a distribution of sizes and lifetimes (and smaller spots have shorter lives, in agreement with sunspot observations, \citealt{Solanki2003}). In contrast to this, the models of \citet{Basri2020} have spots with a single well-defined lifetime. Spot areas grow linearly to a fixed maximum size over half the lifetime then decay symmetrically in these latter models. We remind the reader that these models have lifetimes in units of rotation period, so the rotation period itself can be set to any arbitrary number and the inferred period will scale accordingly. Here the period was arbitrarily set to 10 days simply for display purposes. For each of the lifetimes tested we applied both period-finding methods to each of the 1000 light curves produced. The histograms of the derived periods for all cases are displayed in Fig.~\ref{lifetimes}. The top panel shows that for the GPS method, models where the spot lifetime was longer than one rotation long did very well in finding the correct period. With $\alpha=0.183$ the mean period found for these models is 9.99 days with a 1$\sigma$ dispersion of less than 10\% (see Table \ref{tbl-1}). The model with spot lifetime of one rotation produced a mean period of 8.7 days at the same $\alpha$. This value of $\alpha$ has been post-selected to match the model period of 10 days for those models with lifetimes greater than 2 rotations. We emphasize that this value is well within the $1\sigma$ range determined from the noise-free SATIRE models (see Sect.~\ref{alpha}).

Interestingly, the GPS-derived period (using the same $\alpha$) for a spot lifetime of 0.5 rotations is a little under two-thirds of the actual rotation period, and for a lifetime of 0.25 it is about one-third of the true period. This demonstrates explicitly that the GPS result is sensitive to the typical duration of a dip in the light curve. As a consequence, $\alpha$ does not depend on the rotation period of the model light curve. When rotation is the cause of the typical dip duration (spot evolution occurs on the rotational timescale or longer) GPS delivers the actual rotation period. But when spot evolution is the primary driver of the duration of the typical dip, GPS responds to that more than to rotationally-driven features. Also of interest is the fact that the GPS dispersion is twice as small (percentage-wise) for the shortest lifetime than for the other cases. This may be because the interaction between rotation and spot evolution is weakest in this case; spot evolution completely dominates. This is most relevant for slowly rotating stars, which are more likely to have spot lifetimes that are smaller fractions of their rotation period. The implication is that some of the GPS-derived periods for such stars could be shorter than the real periods. This analysis suggests that for spot lifetimes of 2 periods and less previous analyses (mostly using Kepler data) may not have determined the correct periods.

The results for the auto-correlation method are quite different. Although the mean derived period is nearly correct for the shortest spot lifetime, that is entirely coincidental given the nearly 40\% 1$\sigma$ dispersion of the individual periods (which range between 1 and 20 days). One can place no confidence in any individual result under such circumstances. The mean derived periods for lifetimes of a half and unit rotation periods are around twice what they should be, with similarly large errors making them equally unusable. At a spot lifetime of twice the rotation period the results for just under half of the light curves begin to converge back to the correct period but the majority of derived periods remain much too high and dispersed. It appears that some of the light curves for lifetime two are periodic enough for auto-correlation to detect it while more are not quite there, just due to random variations in the spot patterns. We were unable to identify specific metrics that distinguish the two groups (the light curves look very similar). The auto-correlation method finally converges on just under the correct result when the spot lifetimes are three or more rotation periods, and the dispersion in the derived periods becomes very tight (significantly tighter than for GPS). It can also be noted that the \citet{Basri2020} models that have lifetimes longer than 2 rotations produce light curves that are much like those that \citet{McQuillan2014} were successful with. That this is true is clear from the strong success of the auto-correlation method on them, and the GPS method indeed found the same period.

\begin{table}
\begin{center}
\caption{Results for Lifetime Models.\label{tbl-1}}
\begin{tabular}{rrrrrrr}
\tableline\tableline
Lifetime & P-GPS & $\sigma$-GPS & \%$\sigma$-GPS & P-ACH & $\sigma$-ACH & \%$\sigma$-ACH \\
\tableline
0.25 & 3.35 & 0.16 & 4.8 & 9.84 & 3.89 & 39.5 \\
0.5 & 6.12 & 0.55 &9.0 & 18.3 & 6.08 & 33.2 \\
1.0 & 8.73 & 0.84 & 9.6 & 25.5 & 7.25 & 28.4 \\
2.0 & 9.91 & 0.85 & 8.6 & 19.9 & 10.42 & 52.4 \\
3.0 & 10.08 & 0.81 & 8.0 & 9.72 & 1.51 & 15.5 \\
4.0 & 9.99 & 0.78 & 7.8 & 9.81 &  0.19 & 1.9 \\

\tableline
\end{tabular}
\end{center}
\tablecomments{P-GPS is the mean rotation period for each of the 1000 trial sets using $\alpha=0.183$. P-ACH is the mean period found with the auto-correlation method. All models were generated with an arbitrarily chosen rotation period of 10 days. }
\end{table}

\section{Summary \& Conclusions}

Solar variability is known to be highly irregular on rotational timescales. In this study, we employed physics-based models of stellar brightness variations as these would be seen by the Kepler telescope for stars with different inclination angles, metallicities, and noise levels. Previously, \citet{Reinhold2021} used the same models to show that measuring accurate rotation periods in such highly irregular light curves is a difficult task by itself, and standard frequency analysis techniques such as auto-correlation functions often fail to detect the correct period.

In this study, we took advantage of the GPS method \citep{GPS_I} and demonstrated that it is a powerful tool for detecting rotation periods in such highly-irregular light curves. In particular, the detection rate of $\approx 40\%$ is much higher than the $\approx 3\%$ of the ACF method using common peak height thresholds as $LPH>0.3$ (see \citealt{Reinhold2021}). We further showed that the GPS method is largely insensitive to different inclinations and metallicities but visibly reacts to noise, pushing the calibration factor $\alpha$ toward larger values.

In summary, the distribution of $\alpha$ is broad which is only partly caused by noise but also stems from the presence of the various manifestations of magnetic activity, i.e. the interplay of dark spots and bright faculae (see Fig.~\ref{alpha_activity}). Owing to the relatively large spread of the calibration factor $\alpha$, the derived rotation periods are accompanied by errors on the order of $\approx 20\%$.

Another direct comparison of the GPS and ACF methods showed that the ACF only works in cases where the spot lifetimes are significantly longer than the rotation periods (2 or more times). By contrast, the GPS method yields proper rotation periods for spot lifetimes longer than one rotation period with moderate relative uncertainties. For more periodic light curves, i.e. likely those with long spot lifetimes, both methods converge and yield consistent results, with ACF periods becoming increasingly accurate with increasing spot lifetime (relative to the rotation period). Therefore, whereas for periodic or quasi-periodic light curves, ACF may be the method of choice (giving somewhat lower errors for the most periodic ones), while the GPS method is clearly superior to the ACF method for finding periods in aperiodic light curves.

All these attributes make the GPS method a promising tool for determining rotation periods of those Kepler stars lacking such a measurement so far. This task will be addressed in a forthcoming publication. In particular, we will apply the GPS method to all Kepler stars with near-solar effective temperatures. This effort will generate the largest sample of solar-like stars available so far. It will serve as test bed for various solar-stellar comparison studies.

\bibliography{references}
\bibliographystyle{aasjournal}
\end{document}